\documentclass[aps,prl,twocolumn,groupedaddress]{revtex4-1}

\usepackage{graphicx}
\usepackage{amsmath,amsfonts,amsthm} 

\usepackage{color}

\begin{document}


\title{The Glassy Solid as a Statistical Ensemble of Crystalline Microstates}


\author{Eric B. Jones} 
\affiliation{Colorado School of Mines, Golden, Colorado 80401, USA}
\affiliation{National Renewable Energy Laboratory, Golden, Colorado 80401, USA}

\author{Vladan Stevanovi\'{c}} 
\email[]{vstevano@mines.edu}
\affiliation{Colorado School of Mines, Golden, Colorado 80401, USA}
\affiliation{National Renewable Energy Laboratory, Golden, Colorado 80401, USA}


\date{\today}

\begin{abstract} 
Motivated by the concept of partial ergodicity, we present an alternative description of covalent and ionic glassy solids as statistical ensembles of crystalline local minima on the potential energy surface. We show analytically that the radial distribution function (RDF) and powder X-ray diffraction (XRD) intensity of ergodic systems can be rigorously formulated as statistical ensemble averages, which we evaluate for amorphous silicon and glassy silica through the first-principles random structure sampling. We show that using structures with unit cells as small as 24 atoms, we are able to accurately replicate the experimental RDF and XRD pattern for amorphous silicon as well as the key structural features for glassy silica, thus supporting the ensemble nature of the glasses and opening the door to fully predictive description without the need for experimental inputs.
\end{abstract}


\maketitle

\begin{figure}
\includegraphics[width=0.9\linewidth]{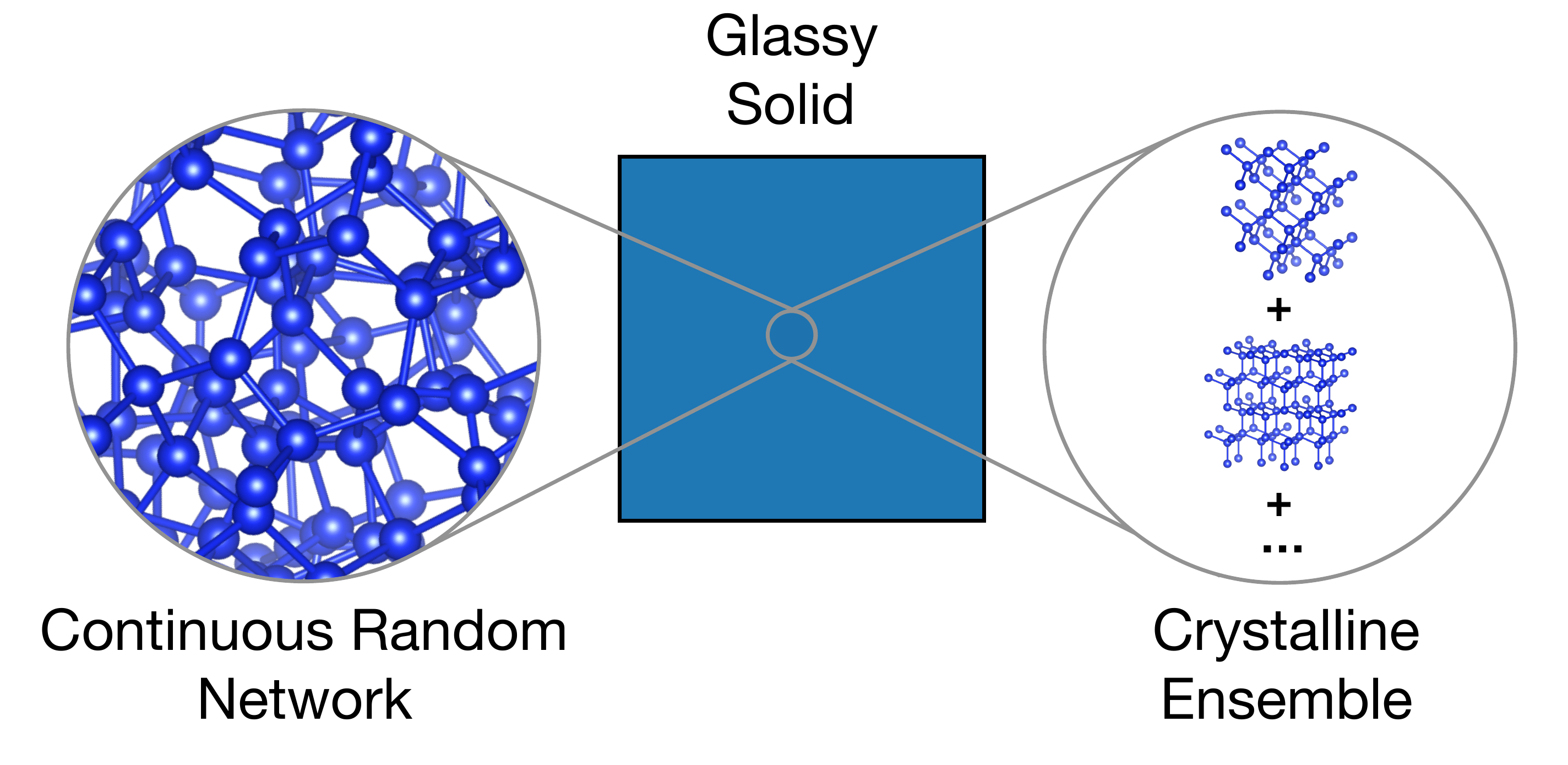}
\caption{Two complementary views of the covalent and ionic glassy solids. The widely accepted continuous random network model describes the glassy state as a single microstate. Meanwhile, our findings speak to a statistical ensemble of crystalline microstates description which features remnant, but reduced, ergodicity from the liquid. \label{fig:concept}}
\end{figure}

Modern approaches to describing the atomic structure of glassy solids are predicated upon the three classical models: ({\it i}) the continuous random network for covalent and ionic systems, ({\it ii}) random close packing for metallic glasses, and ({\it iii}) the random coil model for polymeric glasses \cite{zallen2008physics}. While the random close packing and random coil models are defined statistically over an ensemble of random packings and random conformations respectively, the continuous random network is usually conceptualized in terms of a single ``optimal'' structural model \cite{pedersen2017optimal}. Depicted in the left-hand side of Fig.~\ref{fig:concept}, the optimal structure is a single microstate that minimizes the total energy subject to a certain bonding topology constraint \cite{tu1998properties}.\\
That the covalent (or ionic) glassy macrostate is also identifiable with an \textit{ensemble} of microstates can be understood by considering the instance in which a glassy state is obtained by supercooling from a liquid. A liquid can be described as a point in $3N$ dimensional configuration space thermally sampling different local potential energy minima and their attraction basins \cite{dorner2018melting, sjostrom2018potential}. Liquids are ergodic. Time averages of liquid properties are equivalent to ensemble averages on the timescales accessible to experiment. Upon supercooling, the resulting glassy state is no longer fully ergodic; not every microstate strictly allowable by energy considerations is accessible over experimental timescales. However, given that the configuration point of the liquid roams freely over the potential energy surface, it has been pointed out that a realistic quenching scenario involves the configuration point becoming kinetically constrained to some smaller region of configuration space, partially breaking the full ergodicity, but otherwise able to thermally sample and locally relax between different basins of attraction in the glassy state \cite{PhysRevLett.121.118001, ozawa2018ideal, mauro2014statistical}.\\
In this letter, we present a general theory for describing the atomic scale structure of glassy states, which is in harmony with the concept of only partially broken ergodicity and which allows for the covalent and ionic glassy solids to be treated in a statistical manner on equal footing with metallic and polymeric glasses.
We proceed to show analytically how important structural descriptors of the glassy state can be calculated as averages within such an ensemble framework. The two quantities we choose to calculate are the radial distribution function (RDF), which captures the local order of the state and the powder diffraction intensity (XRD), which captures the long-range order of the state. For our theoretical development we make no assumptions regarding the structural periodicity of the ensemble microstates. However, we will find that evaluating the RDF and XRD expectation values for structural ensembles with a periodic unit cell as small as 24 atoms reproduces experimentally measured RDF and XRD patterns remarkably well for amorphous silicon (a-Si). This result therefore establishes the ``crystalline ensemble'', shown on the right-hand side of Fig.~\ref{fig:concept} as a valid complementary view of the glassy state. Similar ideas have been introduced previously by Curtarolo and co-workers who proposed a spectral descriptor for predicting glass-forming metallic alloys based on enthalpy distribution of ordered structures: the more dissimilar the structures with similar energies, the higher is the probability of formation of the amorphous state  \cite{curtarolo_NM:2016}. \\
%
%
\textit{Ensemble Construction} -- In order to show the decomposition of the glassy radial distribution function into a thermal average, we begin with the expression for the two-point density function for our glassy solid in the isothermal-isobaric $(N,p,T)$ ensemble \cite{ziman1979models}. We choose this ensemble because it is under constant number of particles, temperature, and pressure (rather than volume) that many practical experiments take place, and which therefore need to accommodate microstates of varying volume. We will hence consider the enthalpy as the relevant thermodynamic potential. Later, we will set $p=0$ for simplicity. However, for the present theoretical development, it is important to keep pressure arbitrary so as to allow for microstates of differing volume, as mentioned above, since the volume of the glassy state should be derivable as an expectation value, not taken as an empirical input. The ensemble expectation value for the two-point density function of an ergodic system reads:
\begin{align} \label{eqn:tp_dens}
n^{(2)}(\mathbf{r}',\mathbf{r}'') &= \langle \sum_{i=1}^N \sum_{j \neq i} \delta(\mathbf{r}_i - \mathbf{r}') \delta(\mathbf{r}_j - \mathbf{r}'') \rangle _{(N,p,T)} = \nonumber \\
&= \frac{1}{\Xi} \int dV \int d\mathbf{r}_1 ... d\mathbf{r}_N e^{-\beta(U(\mathbf{r}_1, ... , \mathbf{r}_N , V) + pV)} \nonumber \\
& \times \sum_{i=1}^N \sum_{j \neq i} \delta(\mathbf{r}_i - \mathbf{r}') \delta(\mathbf{r}_j - \mathbf{r}'') , 
\end{align}
where $\Xi$ is the partition function, $U$ potential energy, and $U+pV$ enthalpy of the system of $N$ particles. The double summation in the integrand runs over all particle pairs, and thus, $n^{(2)}(\mathbf{r}',\mathbf{r}'')$ describes the probability that any two particles occupy simultaneously the positions $\mathbf{r}'$ and $\mathbf{r}''$. Next, we split the configurational integrals in eq.~\eqref{eqn:tp_dens} into a sum of integrals over the basins of attraction $B_{\alpha}$ of various potential enthalpy minima (labelled $\alpha$). In this summation we multiply and divide by the intra-basin partition function $\Xi_{\alpha} \equiv  \int_{B_{\alpha}} dV d\mathbf{r}_1 ... d\mathbf{r}_N e^{-\beta(U(\mathbf{r}_1, ... , \mathbf{r}_N , V) + pV)} $.  This allows us to express the two-point density function exactly as a weighted sum over two-point density functions constrained to reside within the basins $\{ B_{\alpha} \}$.
\begin{align} \label{eqn:tp_sum}
n^{(2)}(\mathbf{r}',\mathbf{r}'') &= \sum_{\alpha} P_{\alpha} \, n^{(2)}_{\alpha}(\mathbf{r}',\mathbf{r}'')
\end{align}
%
%
The $P_{\alpha} = \Xi_{\alpha} / \Xi $ in eq.~\eqref{eqn:tp_sum} are the ensemble \emph{probabilities} of individual local minima, which we showed previously to correlate with the experimental realizability of metastable polymorphs~\cite{stevanovic2016sampling,jones2017polymorphism}. To evaluate these probabilities we will assume the ``flat basin approximation'', {\it i.e.}, that the potential enthalpy adopts a square well shape in every basin of attraction. Under this approximation $P_{\alpha} \approx f_{\alpha} \exp(-\frac{h_{\alpha}}{\tau})/\Xi$, where $h_{\alpha}$ is the enthalpy per particle at the local minimum of the basin $B_{\alpha}$, $\tau = k_B T / N$ is an effective (scaled) temperature, and $f_{\alpha}$ is the hypervolume of the basin $B_{\alpha}$. We also showed the $f_{\alpha}$ can be estimated from the relative frequency of occurrence of a local minimum ${\alpha}$ in the first-principles random structure sampling (explained below). Meanwhile, $n^{(2)}_{\alpha}(\mathbf{r}',\mathbf{r}'')$ is the intra-basin expectation value of the two-point density operator. \\
We will approximate the intra-basin average by the two-point density function found at the minimum of the basin $B_{\alpha}$. Given the relationship between the two-point density function and the two-point correlation function $n^{(2)}(\mathbf{r}',\mathbf{r}'')=\langle n \rangle^2 g^{(2)}(\mathbf{r}',\mathbf{r}'') $ with $\langle n \rangle$ the ensemble averaged global number density, we express $g^{(2)}(\mathbf{r}',\mathbf{r}'')$ also as a weighted sum using eq.~\eqref{eqn:tp_sum}
\begin{align}
 g^{(2)}(\mathbf{r}',\mathbf{r}'') &= \sum_{\alpha} \bigg(\frac{n_{\alpha}}{\langle n \rangle}\bigg)^2 P_{\alpha} \, g^{(2)}_{\alpha}(\mathbf{r}',\mathbf{r}'').
\end{align}
$n_{\alpha}$ is the global number density of the minimum of the basin $B_{\alpha}$. In order to derive the radial distribution function, we perform three more steps. First, make the coordinate transformation to a ``displacement'' coordinate $\mathbf{r}=\mathbf{r}' - \mathbf{r}''$ and a ``center of mass'' coordinate, $\mathbf{R} = \frac{1}{2}(\mathbf{r}' + \mathbf{r}'')$. Second, volume average $g^{(2)}(\mathbf{r},\mathbf{R})$ over the center of mass coordinate. And third, average the resulting pair correlation function $g^{(2)}(\mathbf{r})$ over the polar coordinates of $\theta_r$ and $\phi_r$. The final expression for the radial distribution function as a weighted sum is:
\begin{align} \label{eq:RDF}
 g^{(2)}(r) &= \sum_{\alpha} \bigg(\frac{n_{\alpha}}{\langle n \rangle}\bigg) P_{\alpha} \, g^{(2)}_{\alpha}(r).
\end{align}
And using the relationship between the structure factor and the pair correlation function $S(\mathbf{q}) = 1 + \langle n \rangle \int_{\langle V \rangle} d\mathbf{r} e^{-i \mathbf{q} \cdot \mathbf{r}} g^{(2)}(\mathbf{r})$, one can show that the powder diffraction intensity $ I(2\theta)$ can be similarly expressed (see supplementary material) \cite{ziman1979models} as:
 \begin{align} \label{eq:XRD}
 I(2\theta) &= \sum_{\alpha} P_{\alpha} \, I_{\alpha}(2\theta).
 \end{align}
%
%
Having formulated the conditions under which the glassy RDF and XRD can be evaluated by proper statistical averages over sets of potential enthalpy minima, we now describe how computationally to obtain an ensemble of such microstates in order to make an explicit evaluation of the RDF and XRD averages.\\
%
\textit{Methods} --  The computational construction of an ensemble of structural microstates and attendant evaluation of ensemble averages utilizes the first-principles random structure sampling to generate the atomic configurations that reside at local minima, calculate their enthalpy $h_{\alpha}$, and assess their associated basin hypervolumes $f_{\alpha}$ relative to one another. We remark that these three steps have been thoroughly detailed in our previous work \cite{stevanovic2016sampling, jones2017polymorphism} so we simply sketch them here.\\
\begin{figure}
\includegraphics[width=\linewidth]{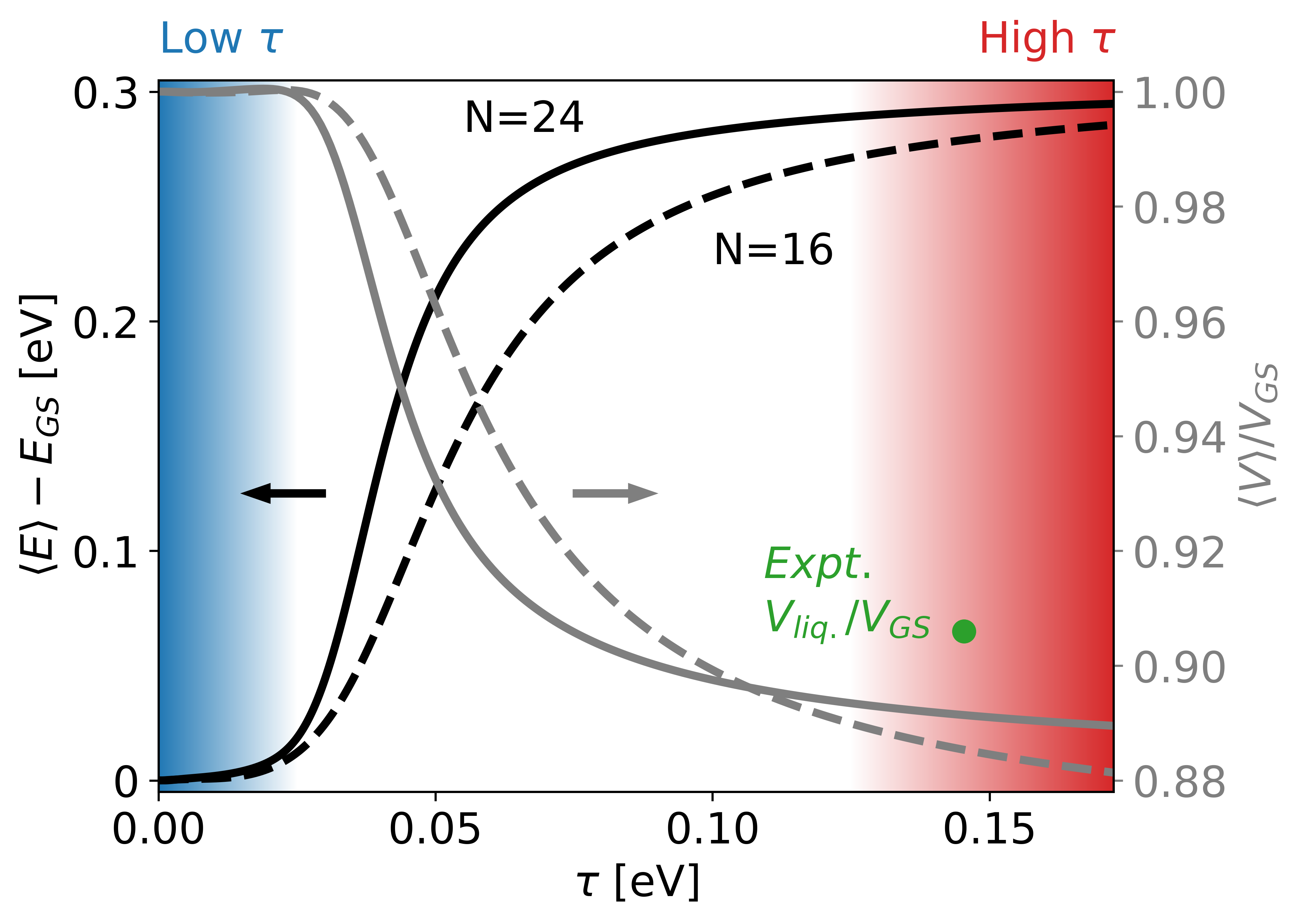}
\caption{ Energy expectation value and fractional volume per atom as a function of the effective temperature $\tau$. A low $\tau$ and high $\tau$ regime are clearly discernible, which correspond to a structurally ordered and disordered state, respectively. \label{fig:energy}}
\end{figure}
The first step is to generate large number of random initial structures with a fixed number of atoms in a unit cell $N$. For Si we will show results for $N=16$ and $24$ while for SiO$_2$ we will cite results just for $N=24$. We find that $N=24$ is sufficient in each case to attain converged RDF and XRD patterns via ensemble averaging. Once $N$ has been chosen, a random unit cell geometry is specified by choosing randomly three lattice constants, $a, b, c$ and the corresponding angles $\alpha, \beta, \gamma$. The unit cell is then populated (quasi)randomly with $N$ number of atoms. In case of Si the population is truly random, while for SiO$_2$ we utilize the random supperlattice construction \cite{stevanovic2016sampling} that biases the structures toward predominant Si-O coordination. 
%
For Si, $15,000$ such random structures were generated with $N=16$ atoms in the unit cell and $10,000$ structures were generated for $N=24$. For SiO$_2$, $4,000$ random structures were generated with $N=24$. Volume, ionic, and cell-shape degrees of freedom were relaxed for all initialized random structures using the VASP implementation of Density Functional Theory (DFT) with the projector augmented wave (PAW) method and and Perdew, Burke, Ernzerhof (PBE) exchange correlation functional \cite{CMS.6.15, PhysRevB.50.17953, PhysRevLett.77.3865}. Calculations were restarted until total energy converged to within $3$ meV$/$atom between successive ionic steps and until residual forces and pressures were below $10^{-4}$ eV $/$ atom and $3$ kbar respectively. Once relaxed, the enthalpy was then evaluated at each minimum using the DFT calculated total energy and volume per atom $h_{\alpha}=E_{\alpha}+pV_{\alpha}$.\\
Finally, note that in the evaluation of an ensemble average, atomic structures that show up more than once will naturally generate a multiplicity $f_{\alpha}$ relative to the total number of structures relaxed due to the additivity of equivalent Boltzmann factors. Therefore, we do not need to sort structures into equivalence classes before taking the average and the relative size of different basins of attraction is naturally included.\\
\begin{figure}
\includegraphics[width=0.8 \linewidth]{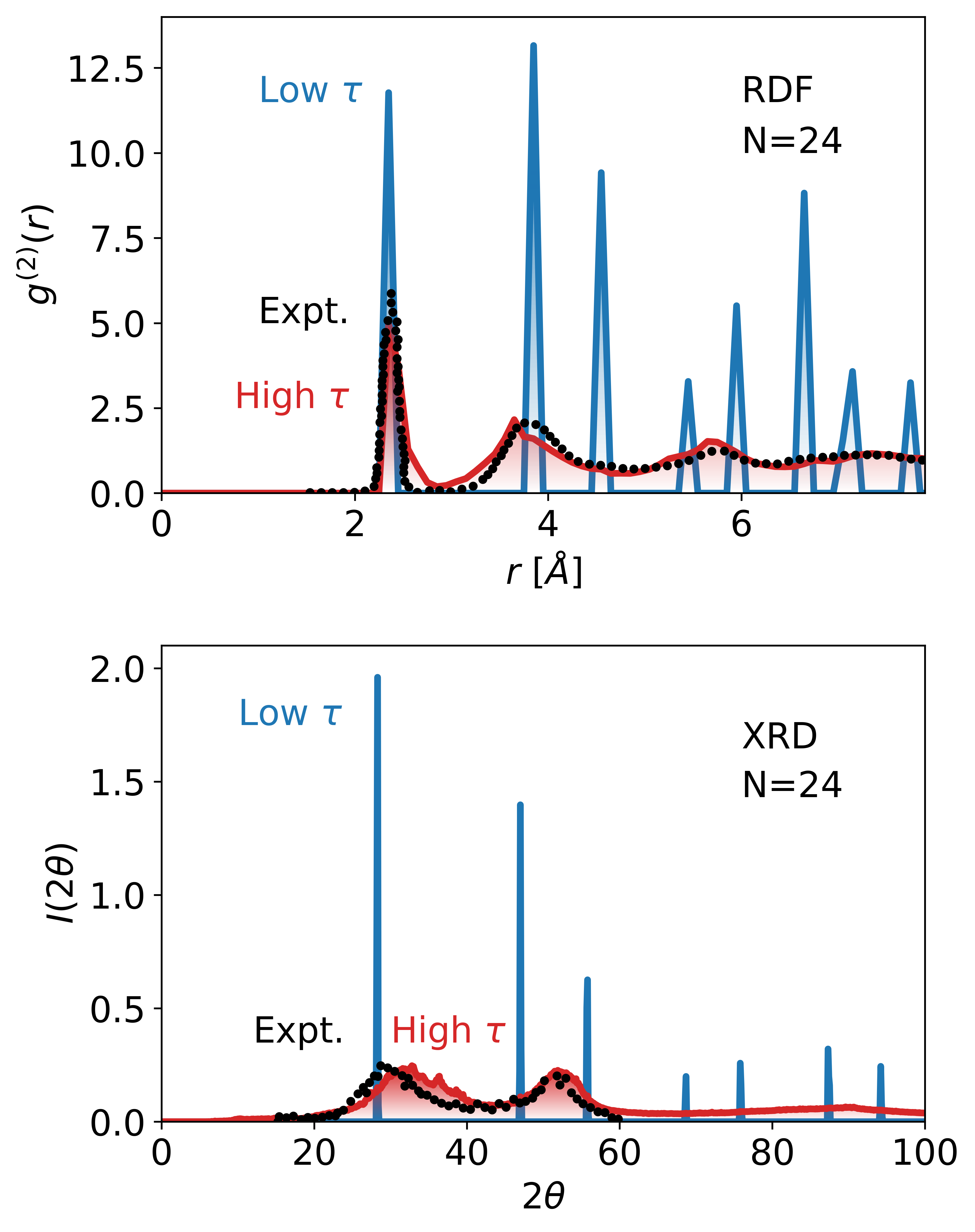}
\caption{ In both the ensemble averaged RDF and XRD patterns, the low $\tau$ regime shows the characteristic peaks of diamond silicon while the high $\tau$ regime displays remarkable agreement with the experimental RDF and XRD for a-Si. \label{fig:exp_v_theory}}
\end{figure}
\textit{Results} -- We note that from hereon, all calculated observable quantities will be cited at $p \approx 0$ GPa in order to be able to reference them against experimental data. The terms ``energy'' and ``enthalpy'' will therefore be synonymous in the following. In order to understand the effective thermodynamics of our ensemble, we plot two important thermodynamic quantities in Fig.~\ref{fig:energy} as a function of the ensemble effective temperature $\tau = k_B T / N$ for elemental Si. The expectation value of the energy per particle relative the diamond Si ground state, $\langle E \rangle - E_{GS}$ is shown in black while the expectation value of the volume as a fraction of the ground state volume, $\langle V \rangle / V_{GS}$ is shown in grey (right axis). Solid (dashed) lines correspond to the sampling results with $N=24$ $(N=16)$ atoms. All curves shown display the sigmoid-type signature of an effective order-disorder phase transition, with increasing unit cell size $N=16 \rightarrow 24$ rendering the character of the phase transition sharper. While the low $\tau$ regime, demarcated by shaded blue region, clearly describes the ground state properties of the system -- zero energy and unit fractional volume -- the high $\tau$ regime (red shading) involves a large density of states with energy roughly clustered around $\sim 0.3$ eV, leading to a high $\tau$ energy expectation, which asymptotes to that value. Meanwhile the fractional volume in the high $\tau$ limit goes to $\sim$ 0.89.\\
Given evidence of an effective order-disorder transition, we now proceed to show that the ground (low $\tau$) state corresponds to diamond silicon (d-Si) and that the high $\tau$ state corresponds to an ideal silicon glass (g-Si). In the top panel of Fig.~\ref{fig:exp_v_theory} we show ensemble radial distribution functions for the low $\tau$ regime (blue) and the high $\tau$ regime (red), averaged according to eq.~\eqref{eq:RDF} with a unit cell of $N=24$ atoms. We also plot the experimental radial distribution function for a-Si prepared by ion implantation (black dots) \cite{laaziri1999high}. The distinct crystalline peaks of the low $\tau$ RDF confirm that our ground state structure indeed corresponds to d-Si \cite{laaziri1999highen}. At high $\tau$ we find that the RDF broadens and overlays the experimental RDF to a rather remarkable degree of accuracy. This agreement in the local ordering of a-Si demonstrates that averaging the local order parameter (RDF) of an ensemble of crystalline arrangements of atoms 
can be described according to the principle of partial, remnant ergodicity.\\
The bottom panel of Fig.~\ref{fig:exp_v_theory} contains powder X-ray diffraction patterns, again for the ensemble low and high $\tau$ regimes averaged according to eq.~\eqref{eq:XRD} and for an experimentally prepared a-Si \cite{hatchard2004situ}. The low $\tau$ peaks again clearly demonstrate the crystallinity of the d-Si ground state \cite{westra2010formation}. Meanwhile, the high $\tau$ averaged XRD faithfully reproduces the two broad experimental humps shown in black. Given that each individual microstate in our ensemble possesses sharp diffraction peaks -- due to a long range order on the periodic 24 atom cell -- the two broad humps in the high $\tau$ regime must result from the thermal superposition destroying the long range order found in all of the constituent microstates. Therefore, the lack of long-range order in a glassy solid can be understood as resulting from an incommensurateness in the long-range order of the system as it slips between its accessible microstates.\\
Having established that the crystalline ensemble correctly reproduces both the short and long range order parameters of a-Si, we turn to a discussion of two other important measured properties of disordered silicon states, the excess enthalpy and the density, in order to elucidate the high $\tau$ state as describing silicon glass. The excess enthalpy of liquid silicon over that of the crystal has been measured to be $\sim$ 0.47 eV/atom \cite{veryatin1965thermodynamic}. Recall that the high $\tau$ state has an excess energy of $\sim$ 0.3 eV. From the perspective that a liquid shares many of the same configurational states with its corresponding glass and differs mostly with respect to some added kinetic energy, the high $\tau$ state sits at an energy value identifiable with glassy states potentially accessible by very fast melt quench. This view is supported by the fact that the fractional volume asymptotes in the high $\tau$ limit to $\sim$ 0.89 at $N=24$, representing a $2\, \%$ error relative to the experimental value of the volume of liquid Si as a fraction of the volume of the diamond Si ($V_{liq.}/V_{GS}\approx0.91$), shown as a green point in Fig.~\ref{fig:energy}. Note that the effective temperature at which the experimental $V_{liq.}/V_{GS}$ is plotted corresponds to the physical melting temperature of diamond silicon (d-Si), $T=1687$ K \cite{ubbelohde1978molten}.\\
We therefore predict that the ideal silicon glass should be more dense than diamond silicon, consistent with the fact that liquid Si is also more dense. The ideal continuous random network (CRN) model also predicts that silicon glass should be more dense since any variation in bond angles from perfect tetrahedral coordination densifies the structure \cite{popescu2006amorphous}. By contrast, a-Si has been measured to be $1.8\,\%$ less dense than d-Si \cite{custer1994density}. This density deficit is attributable to a large concentration of coordination defects in the glassy structure due to the ion implantation \cite{brodsky1972densities, williamson1995nanostructure, laaziri1999high, pedersen2017optimal}, which also lower the excess energy of a-Si to around $\sim0.07 - 0.15$ eV/atom as measured by differential scanning calorimetry (DSC) \cite{roorda1989calorimetric, kail2011configurational}. A quenching method that is fast enough to produce the true glass transition should, according to our results and the discussion from Ref.~\cite{popescu2006amorphous}, create the high $\tau$ silicon glass.\\
\begin{figure}
\includegraphics[width=0.8 \linewidth]{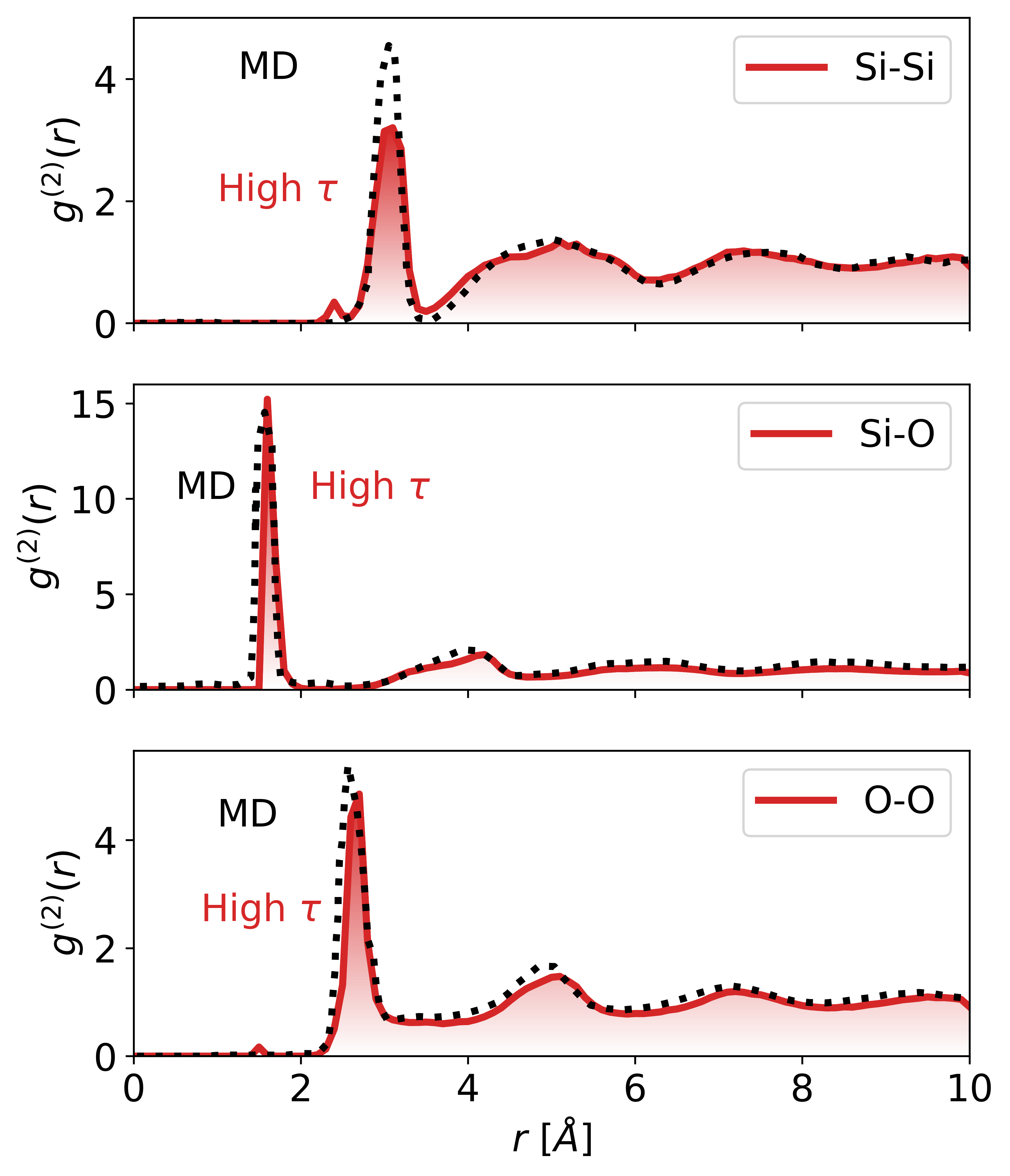}
\caption{  Partial radial distribution functions for the components of vitreous SiO$_2$.  At high $\tau$, the RDF expectation values agree well with previously calculated molecular dynamics results from Hoang \cite{hoang2007molecular}. \label{fig:SiO2}}
\end{figure}
Finally, in order to demonstrate that our results are generally applicable, we show three partial radial distribution functions for vitreous SiO$_2$ in Fig.~\ref{fig:SiO2}. The red curve in each panel corresponds to the high $\tau$ state while the dotted black line represents the RDF from a structural model of amorphous silica (a-SiO$_2$) obtained by a molecular dynamics calculation by Hoang on a 3000 atom unit cell \cite{hoang2007molecular}. Here again, the high degree of agreement between the partial RDFs calculated as a single microstate on a large supercell (black) and our high $\tau$ RDFs calculated as ensemble averages of 24 atom crystalline microstates (red) establishes the mutual compatibility of these viewpoints in describing glassy structure. Furthermore, the first coordination shells of Si with O and O with Si are computed to be $\sim4.01$ and $\sim2.01$ respectively, consistent with known experimental measurements~\cite{prescher2017beyond}. And finally, the small discrepancy in peak heights in the Si-Si and O-O components of Fig.~\ref{fig:SiO2} are a result of a splitting of the first coordination shells and attendant creation of small peaks below the hard-core radius due to a small set of crystalline microstates with substitutional disorder between Si and O sites. This splitting is expected to wash out in a larger sampling of the potential energy surface.\\
\textit{Conclusion} -- We have shown that both the short and long range order of covalent and ionic glassy solids can be accounted for by taking thermal averages over an ensemble of crystalline microstates, thus validating the concept of remnant partial ergodicity in these systems. We have also laid out the framework and the approximations within which the ensemble picture rigorously follows from the statistical treatment of fully ergodic systems. The partial ergodicity is taken into account by considering the potential enthalpy of the system and including only low-energy local minima obtained through the first-principles random structure sampling. Such an approach is consistent with the principle of remnant ergodicity within a region of configuration space and affords a fully first principles (no fitting to experiments) accounting of the structural features of glassy Si and SiO$_2$. At the same time,  the crystalline ensemble model opens the door to calculating functional properties of glasses by averaging over the contributions to those properties carried by each microstate in the ensemble.
%
%
\begin{acknowledgments}
V. S. acknowledges inspiring and fruitful discussions with Prof. Artem Oganov from Skolkovo Institute of Science and Technology. This work was supported as part of the Center for the Next Generation of Materials by Design, an Energy Frontier Research Center funded by the U.S. Department of Energy, Office of Science, Basic Energy Sciences. The research was performed using computational resources sponsored by the Department of Energy's Office of Energy Efficiency and Renewable Energy and located at the National Renewable Energy Laboratory.
\end{acknowledgments}
%

%

\end{document}